\documentstyle[epsf]{lamuphys}
\makeatletter
\let\chapter\hid@chapter
\makeatother

\begin{document}
\pagenumbering{arabic}
\title{Jets and QSO Spectra}

\author{Beverley J.\,Wills and M. S.\,Brotherton}

\institute{Department of Astronomy, University of Texas at Austin, Texas,
78712}

\maketitle

\begin{abstract}

QSOs' emission lines arise from highest velocity ($\sim 10^4$\ km s$^{-1}$),
dense gas within
$\sim$\ 0.1 parsec of the central engine, out to low-velocity, low-density gas
at great
distances from the host galaxy.  In radio-loud QSOs there are clear indications
that the
distribution and kinematics of emission-line gas are related to the symmetry
axis of the
central engine, as defined by the radio jet.
These jets originate at nuclear distances $< 0.1$\ pc --- similar to the
highest-velocity
emission line gas.
There are two ways we can investigate the
different environments of radio-loud and radio-quiet QSOs, i.e., those with and
without
powerful radio jets.  One is to look for optical-UV spectroscopic differences
between
radio-loud and radio-quiet QSOs.  The other is to investigate dependences of
spectroscopic
properties on properties of the powerful jets in radio-loud QSOs.
Here we summarize the spectroscopic differences between the two classes, and
present known
dependences of spectra on radio core-dominance, which we interpret as
dependences on the
angle of the central engine to the line-of-sight.
We speculate on what some of the differences may mean.

\end{abstract}
\section{Introduction}

The optical-ultraviolet emission-line profiles and intensity ratios are
remarkably similar
for radio-loud QSOs (RLQs) and radio-quiet QSOs\footnote
{`QSO' refers to all luminous AGN (L$\ga 10^{11}$\,L$_{\sun}$, H$_0 = 100$\,
km s$^{-1}$\,Mpc$^{-1}$.  A radio-loud QSO is one having
F$_{\rm5GHz}$/F$_{4400} \ga 10$,
where F is the rest frame flux-density in mJy.  Such strong radio emission is
assumed
to indicate powerful radio jets.  For a short `course' on emission lines, see
Netzer (1990),
and reviews to appear in the Proceedings of IAU Colloquium No. 159, {\em
`Emission Lines in
Active Galaxies'} (Shanghai, 1996).} (RQQs),
for a wide range of luminosities (Baldwin et al. 1995).
The broad emission lines arise predominantly by photoionization of the
broad-line
region (BLR), a region of dense gas within $\sim$\ 0.1 parsec of the central
engine.  Lower
velocity ($\sim 500$\,km s$^{-1}$) gas of the narrow-line region (NLR) arises
at distances
from parsecs to kiloparsecs.  The BLR and NLR are not single, homogeneous
regions: the
highest-velocity, highest-density gas ($\sim 10^{12 - 13}$\,cm$^{-3}$) --- the
very broad
line region (VBLR) --- occurs closest to the central engine and produces the
broadest
emission lines ($\sim 10^4$\,km s$^{-1}$).  These are typically blueshifted by
$\sim$1000
km s$^{-1}$\ with respect to the systemic (NLR) redshift.  Gas of the ILR ---
intermediate in
velocity, density, and distance, between the VBLR and NLR --- produces profiles
of $\sim
2000$\,km s$^{-1}$\ width.
Much of the diversity in line ratios and profiles can be reproduced by a
combination of
spectra from the VBLR, ILR and NLR (Brotherton et al. 1994b).

The powerful, synchrotron-emitting radio jets arise within VBLR
distances, and often extend hundreds of Kpc from the nucleus.
Thus, differences in the optical-ultraviolet spectra of the BLR between the
RLQs
with powerful jets, and the
RQQs of 1000 -- 10,000 less radio luminosity, should give clues to differences
in material that fuels or is expelled by the central engine, and how this might
be related
to the formation and collimation of the jets.
Differences in the optical-ultraviolet spectra of the lower-speed gas might
reflect jet
interaction with material in the inner regions of the host galaxy, up to many
Kpc distant
--- by means of entrainment, shock excitation, and photoionization by beamed
high-energy
synchrotron emission.

The spin axis of the central engine is indicated by the position angle of the
base of the
jet, and by the radio core dominance that indicates the angle of the jet to the
line-of-sight (see the chapter, `Accretion and Jet Power').
Thus dependence of spectral properties on core dominance
provides a statistical way to investigate axisymmetric structure of the gaseous
environment in RLQs, necessary for the interpretation of radio-loud --
radio-quiet
differences.

First we summarize spectroscopic differences between RLQs and RQQs, then
summarize some spectroscopic dependences on radio core-dominance.
We suggest a picture for future testing, which is consistent with the present
observations.

\section{Radio-loud -- Radio-quiet Spectroscopic Differences}

Despite the great similarity of RLQ and RQQ spectra, differences are being
recognized as a result of improved data over a wide spectral range, for
appropriate samples
of QSOs.
We present results that derive from our own and other investigations, using
ground-based
spectrophotometry in the optical-infrared, and ultraviolet spectroscopy with
the Hubble
Space Telescope, emphasizing those that
are most simply interpreted without detailed statistical analysis.  Even the
strongest
relations involve more than two variables, and multivariate analyses of
carefully defined
samples are needed.  Some of this work is in progress.

\subsection{The Fe\,II -- [O\,III] Relation}

One of the strongest correlations is the inverse correlation between the
strength of
blended optical Fe\,II emission (between H$\gamma$\ and H$\beta$, and 5150 --
5300 \AA),
and the strength of [O\,III]\,$\lambda$5007 --- the strongest optical line from
the NLR.  This
was nicely presented by Boroson \& Green (BG, 1992) for Seyfert 1 galaxies and
QSOs from the PG survey --- a predominantly radio-quiet sample.  Fig. 1
illustrates typical
spectra of the H$\beta$\ region, and in Fig. 2, the inverse relation is
shown by comparing equivalent widths, including BG data for RQQs (open
symbols), but adding data for RLQs from the investigation by Brotherton (1995;
filled
circles).  Two blazars were excluded, because of their strong, variable
synchrotron
continua.  The samples cover comparable optical luminosities.
There is a striking difference between the RLQs and RQQs: the RLQs
appear clustered towards the weak-Fe\,II, strong-[O\,III] end of the relation.

\begin{figure}
\vspace{7.4cm}
\caption[]{The H$\beta$--[O\,III]\,$\lambda$5007 region in a typical radio-loud
and typical radio-quiet QSO.  For the radio-quiet QSO, Fe\,II blends contribute
at all
wavelengths in this region, and [O\,III]\,$\lambda$5007 is barely visible.  For
the ra
dio-loud
QSO Fe\,II blends are weaker, and stronger NLR emission of
[O\,III]\,$\lambda$5007 and

[O\,III]\,$\lambda$4363 is present.
}
\end{figure}

The [O\,III] is representative of NLR emission.  The strong optical
Fe\,II emission must arise in regions of high density and high optical depth to
the ionizing
continuum, and the
line widths of blended Fe\,II appear to be similar to other broad lines,
suggesting an
origin in the BLR.  However, problems in explaining the great strength of
Fe\,II optical
emission
have led to suggestions of a different source of excitation from the standard
ultraviolet
lines (Ly\,$\alpha$, C\,IV\,$\lambda$1549, C\,III]\,$\lambda$1909).  This
problem is
probably related to the general `energy budget' problem, where the observed
ultraviolet
continuum does not appear strong enough, in general, to produce the strengths
of low
ionization lines .  Assuming that the Fe\,II emission does in fact arise in the
BLR, the
inverse correlation can be explained by object-to-object differences in
covering of the
ionizing continuum by
the BLR gas.  The greater the BLR covering, the greater is the shielding of the
more distant
NLR from ionizing photons --- a possibility also considered, but fleetingly, by
BG.

\begin{figure}
\vspace{8.4cm}
\caption{(left) The equivalent width of [O\,III]\,$\lambda$5007 vs. the
equivalent
width of optical Fe\,II.  Open circles are for RQQs from BG,
filled circles for RLQs (Brotherton 1995), open triangles for high-ionization
BAL
QSOs, and stars for low-ionization BAL QSOs.  Upper limits for
EW[Fe\,II] are 1.5\,$\sigma$;
for EW\,[O\,III], 3\,$\sigma$.
}
\caption{(right) The EW of [O\,III]\,$\lambda$5007 vs. the width (FWHM) of
the C\,III]\,$\lambda$1909 feature.  Data are from Brotherton (1996). The
symbols are
the same
as for Fig. 2.}
\end{figure}

\subsection{Broad Absorption Lines and Associated Absorption}

Broad absorption lines (BALs) --- troughs of absorption extending between
$\sim$5,000 km
$^{-1}$\ and 25,000 km s$^{-1}$\  blueward of the corresponding emission line
peak ---
are recognizable
in $\sim$10\% of QSOs, but have never been seen in RLQs (Stocke et al. 1992).
Associated
absorption is narrow, intrinsic absorption occurring near (within a few hundred
km s$^{-1}$) or
blueward of the emission line peak.  Associated absorption is common in BAL
QSOs and occurs
in other RQQs.  It is also common in RLQs --- probably occurring more often in
lobe-dominated RLQs.
Two classes of BAL QSO are distinguished --- high-ionization BAL QSOs with
C\,IV, N\,V and
Ly\,$\alpha$\ BALs, and low-ionization BAL QSOs, which show, in addition to the
high-ionization BALs,  BALs of Mg\,II,
Al\,III, and probably Fe\,II and Na\,I.
Under current debate is the hypothesis that BAL QSOs and other RQQs are the
same --- the observed difference depending on whether or not intrinsic broad
absorbers
happen to lie along our line-of-sight to the nucleus.  This hypothesis receives
strong
support from the general
similarity of their broad-emission-line ultraviolet spectra (Weymann et al.
1991).

However, low-ionization BAL QSOs do show some other spectral differences.
One of these is illustrated in Fig. 2 (stars and open triangles represent low-
and
high-ionization BAL QSOs, including one [O\,III]\,$\lambda$5007 upper limit)
---
nearly all the low-ionization BAL QSOs have very weak [O\,III] and are
super-Fe\,II emitters.
Nearly all known low-ionization  BAL QSOs show significant scattering
polarization and
significantly reddened continua.  The PG QSOs are selected by UV-excess and are
therefore
biased
against being dust-reddened; however, many QSOs selected by high infrared
luminosity (the
IRAS ultraluminous QSOs) show high scattering polarization and reddened
continua --
and a significant fraction show low-ionization BALs (based on small numbers, as
yet), and/or
belong to the rare class of super-Fe\,II emitters.
The strong Fe\,II is also significantly associated with narrow H$\beta$\ from
the BLR, and
softer X-ray spectra (0.3 -- 2\,kev) (Laor et al. 1994, for PG QSOs; Boller et
al. 1995,
\& Grupe et al. 1995, for soft X-ray ROSAT AGN).  The
polarization, reddening, and BALs can be understood as line-of-sight effects;
the Fe\,II
emission, H$\beta$\ width and X-ray slope are not as simply interpreted.  We
therefore do not
suggest
that the entire [O\,III]--Fe\,II anticorrelation is a line-of-sight effect;
there is probably
more than one physical cause.

Recently BALs and narrower associated absorption have been linked with both
warm and cool
X-ray edge absorption, placing the absorbing material within or just beyond BLR
distances
from the nucleus (e.g. Mathur et al. 1995).

\subsection{Stronger ILR emission in Quasars?}

Fig. 2 shows that radio-loud QSOs tend to have stronger emission from the NLR
compared with
the BLR.
The cause may be related to the alignment between jets and extended narrow-line
emission,
resolved on scales of kpc, and between radio power and the strength and width
of [O\,III]
emission in radio galaxies (e.g. Baum \& Heckman 1989; Whittle 1992).
Narrower, but
stronger, BLR lines of C\,IV\,$\lambda$1549 and C\,III]\,$\lambda$1909 in RLQs
compared
with RQQs suggest a greater contribution from ILR gas (Brotherton et al. 1994a;
Francis et al. 1993), and this is further supported by a link between the ILR
and NLR gas,
being investigated by
Brotherton (1996), and illustrated by the inverse correlation between [O\,III]
strength and
C\,III]\,$\lambda$1909 width  (Fig. 3).

Hypotheses to explain the enhanced emission from lower-velocity gas in RLQs
include
jet-shocked gas, jet-induced star-formation, and ionization by beamed
ultraviolet
synchrotron emission.

\subsection{BLR Line Asymmetries}

There are significant statistical differences in line profiles:
\begin{description}
\item[$\bullet$] C\,III]\,$\lambda$1909 and C\,IV\,$\lambda$1549 are narrower
in RLQs than in
RQQs.
\item[$\bullet$] The C\,IV line generally has stronger red than blue wings in
RLQs, but
the blue wing is often stronger than the red in RQQs (Wills et al. 1993, 1995).
\item[$\bullet$] The H$\beta$ broad line also often has stronger red wings in
RLQs, with
the RQQs showing similar frequency of red and blue asymmetries (e.g.
Sulentic 1989, Corbin 1993; BG92).
\end{description}

Systematic line asymmetries imply radially flowing gas with obscuration.  The
most likely
obscuration related to the emission-line region seems to be either dust within
the
unilluminated backside of gas clouds, or an emission region within an obscuring
torus.
In these cases the above line asymmetries suggest that RQQs, and perhaps all
QSOs,
 have greatest inflow in
the innermost BLR, but
that in RLQs, additional outflow occurs, often producing redshifted line wings.

Another likely contributor to some line asymmetries is blending by other
emission lines,
for example, Fe\,II in the red wing of C\,IV\,$\lambda$1549.  In some QSOs with
strong
Fe\,II emission or BALs, the $\lambda$1909 feature is dominated by blended
Fe\,III emission
(Hartig \& Baldwin 1986; Baldwin et al. 1996), and may well account for the
larger width of
this `C\,III]' feature in RQQs.

\section{Radio Core-dominance Relationships for RLQs}

The previous chapter (Accretion and Jet Power) presented evidence that, for
RLQs,
core-dominance  is a
measure of orientation of the jet axis to the line-of-sight, and we retain that
interpretation here.

\subsection{Line Widths \& Asymmetries}

The width of the broad H$\beta$\ line (FWHM) is inversely correlated with
core-dominance
and we have suggested that this is the result of viewing predominantly planar
motions
that are perpendicular to the jet axis: in core-dominated RLQs the radial
velocities are
smaller, in lobe-dominated RLQs, up to $\sim$8,000 km $^{-1}$, and in
broad-line radio
galaxies where we may be viewing the central QSO at even higher inclination,
widths
up to $\sim$20,000 km $^{-1}$ are found (Wills \& Browne 1986; Wills \&
Brotherton 1995;
see also the previous chapter, {\it Accretion and Jet Power}).  This dependence
is weaker for the
C\,IV\,$\lambda$1549 line, which has a greater contribution from ILR gas than
does H$\beta$\
(Brotherton 1995).

Core-dominated RLQs have stronger red wings for C\,IV\,$\lambda$1549 (Wills et
al. 1995;
see Barthel, Tytler \& Thomson 1990).  For a higher-redshift sample, Corbin
(1991) found
a trend in the opposite sense, but this was a marginal result.  On the other
hand,
 for H$\beta$, it is the lobe-dominated
RLQs that often show strong red wings (e.g. Fig. 1; BG; Brotherton 1996).
For examples of H$\beta$\ line asymmetries, see Fig. 6 in BG.
Again, line asymmetries imply radial flow plus obscuration, so these
correlations
suggest axial flow of high-ionization gas, and radial flow of low-ionization
gas in a plane
perpendicular to the jet axis.

\subsection{Associated Absorption and Reddening}

Excluding from consideration those core-dominant RLQs (blazars) with steep,
beamed IR --
ultraviolet synchrotron continua, there
is a strong trend for the most lobe-dominant sources to show steeper
optical-ultraviolet
continua.  This has been seen for the 3CR sample (see the lobe-dominated
sources in Smith
\& Spinrad 1980), especially when the broad-lined radio galaxies are included,
and is shown
more quantitatively for the 408 MHz Molonglo Quasar Sample (Baker \& Hunstead
1995).
Baker \& Hunstead find the same reddening trend in the Balmer decrements.

There is a corresponding trend for increased numbers of associated absorption
systems in
lobe-dominated RLQs (Aldcroft et al. 1994, Wills et al. 1995).

All these trends suggest increasing concentrations of low-ionization material
at the largest
angles to the jet axis, some of which must be close to the active nucleus (see
X-ray
absorption results, \S 2.2).

\subsection{Fe\,II and [O\,III] Emission}

The larger Fe\,II(optical)/[O\,III] and H$\beta$/[O\,III] ratios are seen in
the most
core-dominated RLQs (Jackson \& Browne 1991), and the ultraviolet `Little Blue
Bump',
composed of blended Fe\,II emission and Balmer continuum, shows the same trend
in the
Molonglo Quasar Sample (Baker \& Hunstead 1995).  Jackson and Browne
interpreted this
trend in terms of axisymmetric Fe\,II line emission.  Perhaps the
Fe\,II-emitting
region occurs near the inner edge of a dusty torus and is therefore shielded
from the
observer more readily
at higher inclination angles than other broad lines produced nearer the center.

The stronger
red wings on the C\,IV line for core-dominated RLQs, mentioned above, could be
low-optical-depth ultraviolet Fe\,II emission, related to the greater strength
of Fe\,II
(optical) in core-dominated RLQs.

\section{Summary and Discussion}

Despite the great similarity between the spectra of RLQs and
RQQs, the following observational differences are statistically very
significant:\hfil\break
In RLQs lower-velocity (NLR, ILR) emission lines are more prominent, and C\,IV
and H$\beta$\ emission lines are asymmetric with stronger red wings.
While there is a wide dispersion in properties, RQQs have, on average, stronger
Fe\,II emission, and $\sim$10\% show strong, broad-absorption troughs of high-
and
low-ionization gas -- a phenomenon that is apparently unique to RQQs.

Two apparently independent relations account for much of the diversity in the
optical-ultraviolet
spectra:\hfil\break
(i) For RLQs, with increasing core-dominance, emission lines are narrower,
Fe\,II emission is
stronger, and C\,IV more often has stronger red wings.  With decreasing
core-dominance,
H$\beta$\ more often has stronger red wings, and the liklihood of associated
absorption and
reddening increases.
\hfil\break
(ii) For RQQs, there is an inverse relation between the strengths of Fe\,II and
[O\,III]\,$\lambda$5007 emission, with the strong-Fe\,II--weak [O\,III] QSOs
being
associated with low-ionization BALs, reddening and polarization.

We note that the properties accounting for the radio-loud -- radio-quiet
differences are those
involved in these two relations.

If we interpret line profiles as velocity profiles (rather than as blended
emission), and
assume dust exists on the unilluminated sides of photoionized gas regions and
in a torus,
then we can interpret the observations as follows:\hfil\break
For RLQs, increasing radio core-dominance means smaller inclination of the
rotation axis
to the line-of-sight.  Kinematics, line emission, and the distribution of dust
and absorbing
gas are axisymmetric.  Broader emission lines
at higher inclinations means greater velocities in a plane perpendicular to the
axis.
Red wings on C\,IV for core-dominant RLQs imply high-ionization axial outflow,
and red
wings on H$\beta$\ for lobe-dominated RLQs imply lower-ionization planar
outflow.  At
higher inclinations (edge-on view), high-density, low-ionization
Fe\,II-emitting gas,
being located at the inner edge of the purported dusty torus,
is partially shielded from view, and the line-of-sight to the nucleus is more
likely to pass
through dust and outflowing associated absorbers.  Perhaps Fe\,II-rich grains,
having the
highest evaporation temperatures, exist at the inner edge of the torus,
producing Fe\,II-rich
gas.

Comparison with the profiles of RQQs suggests that the outflow of emission-line
gas
is unique to RLQs.  Blueshifted VBLR emission seen in C\,IV in RQQ may be
common
to all QSOs (in RLQs, this is difficult to determine because of the presence of
stronger red
wings), indicating higher-ionization, higher-velocity inflowing gas closer to
the central
engine.  This inflow may be related to accretion.

It has not been determined whether distribution of nuclear gas in luminous RQQs
is
symmetric about a rotation axis,
although the existence of several examples of alignment between weak radio
structure and
ionization cones in some quite luminous Seyfert galaxies suggests that this
might be the
case (NGC 1068, Antonucci 1993).
BAL material does in some instances cover less than 4$\pi$\,sr, so the
appearance of BAL
QSOs must depend on orientation.
It would be important to understand whether the distribution of BAL gas is
related to any
structure axis.

What underlies the Fe\,II -- [O\,III] anti-correlation?  We suggested that
decreasing covering
of the ionizing continuum by dense, thick, Fe\,II-emitting gas could
simultaneously
reduce Fe\,II emission and allow escaping photons to ionize lower-density,
more-distant gas,
increasing [O\,III] emission.  The characteristics of
line-of-sight dusty material and low-ionization high-velocity outflowing BAL
gas dominate
increasingly with increasing Fe\,II and decreasing NLR ([O\,III]) strengths.
If all RQQs
have BALs then orientation is important in determining BAL characteristics.
If this relation were
one of orientation, then the different locations of RLQs and RQQs in Fig. 2
would
require them to be intrinsically different.
In the next chapter we suggest ways in which the inner-galaxy environment
may explain the observed radio-loud -- radio-quiet differences.

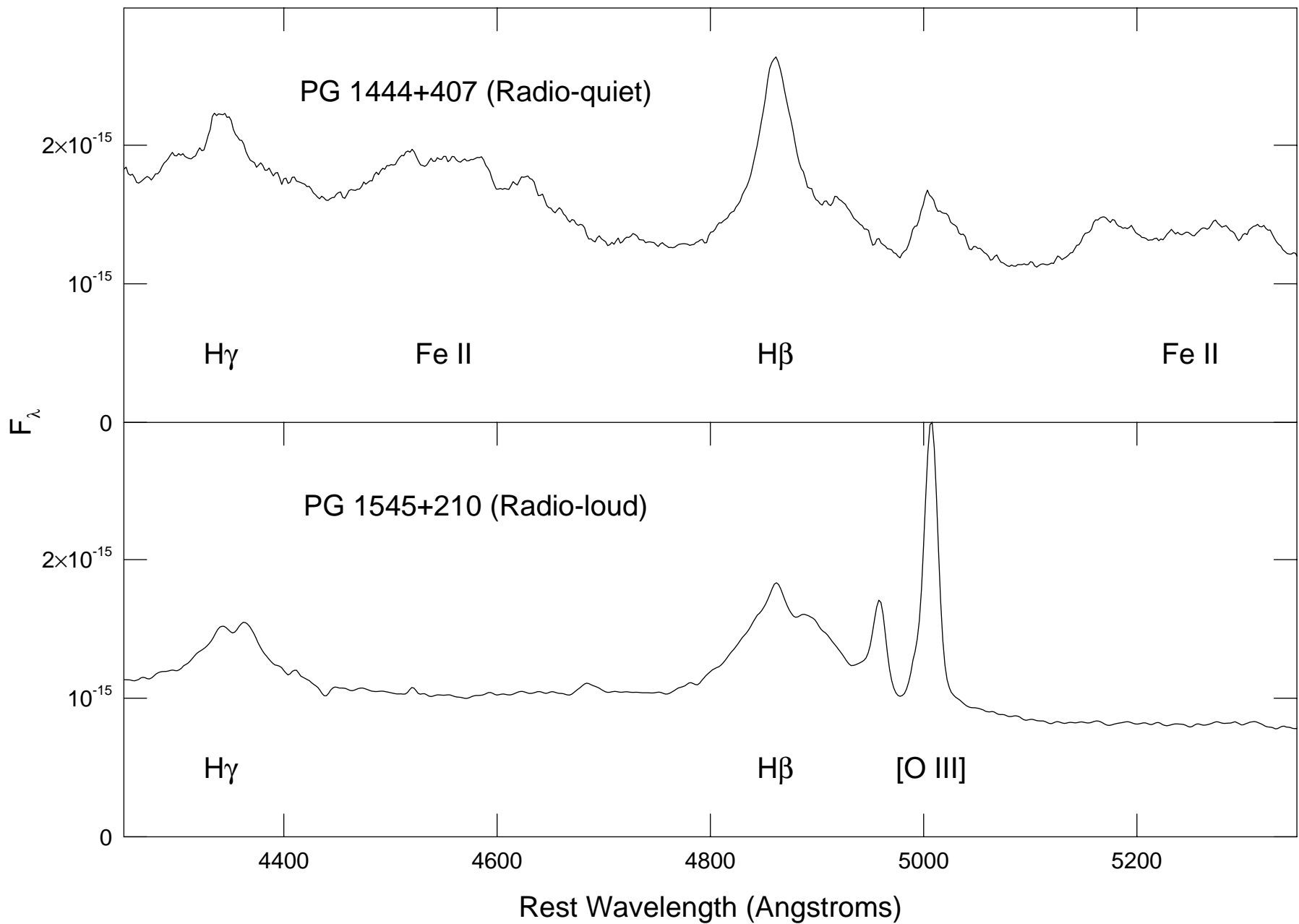
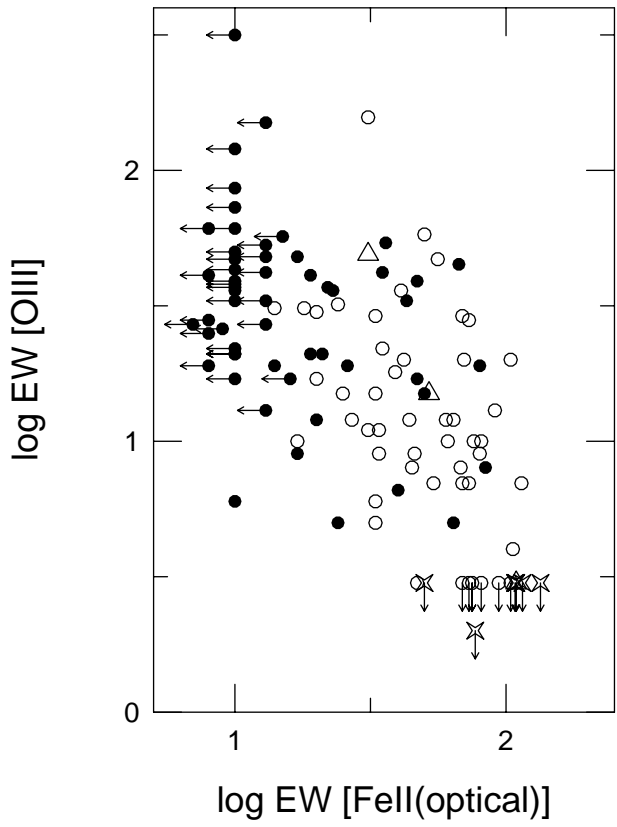
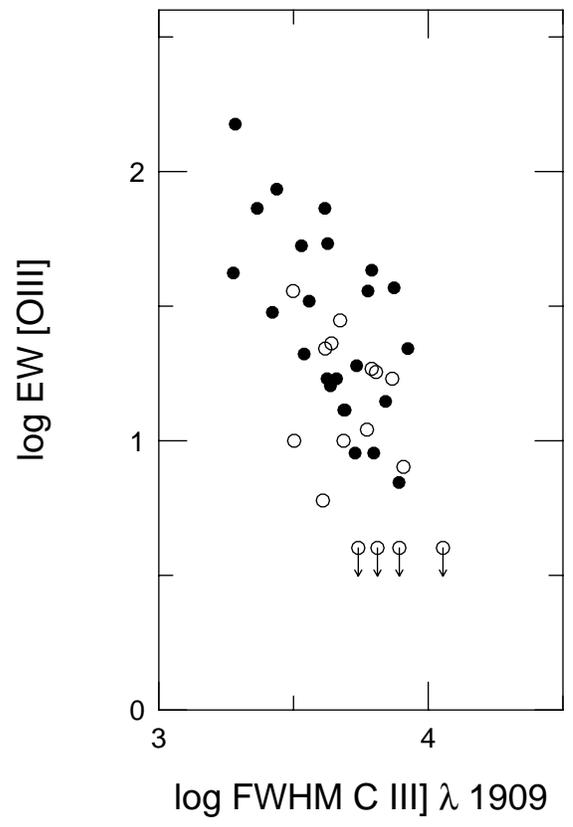
\end{document}